\documentclass[prb,floats,twocolumn]{revtex4}
\usepackage{graphicx}
\begin{document}

\title {Hole dynamics in canted antiferromagnets }

\author{I. J. Hamad$^1$, L. O. Manuel$^1$, G. Martinez$^2$, and A. E. Trumper$^1$}

\affiliation {$^1$Instituto de F\'{\i}sica Rosario (CONICET) and
Universidad Nacional de Rosario,
Boulevard 27 de Febrero 210 bis, (2000) Rosario, Argentina\\
$^2$ Instituto de F\'isica, UFRGS, 91501-970 Porto Alegre, RS,
Brasil
}

\vspace{4in }

\date{\today}
\begin{abstract}
We have analyzed the dynamics of a single hole doped  in a canted antiferromagnet using the $t-J$ model. Within the self consistent Born approximation we have found that the hole propagates at two different energy scales along the antiferromagnetic and the ferromagnetic components of the canted order, respectively. While the many body quasiparticle excitation has its origin in the coherent coupling of the hole with the magnon excitations of the antiferromagnetic component, the ferromagnetic component gives rise to a free like hole motion at higher energies. We have found a non-trivial  behavior of the hole spectral function with the canting angle $\theta$. In particular, in the strong coupling regime, the quasiparticle weight strongly depends  on the momenta,  vanishing inside the magnetic Brillouin zone for $\theta \gtrsim 60^{\circ}$.

\end{abstract}
\maketitle

\section{introduction}

 The underlying physics behind the doped Mott insulators is essential to elucidate the 
 mechanism that leads to the high temperature 
superconductivity. The major challenge is to explain, microscopically, how an antiferromagnetic (AF)
insulator evolves into a superconductor\cite{anderson87}. In this sense, angle resolved photoemission spectroscopy (ARPES) is one of the most refined 
experiments that gives access to the one particle excitations of doped Mott insulators\cite{damascelli03}. Although 
the photoemission spectra of the cuprates have been well characterized by ARPES experiments as a function of doping, there  is still no consensus about the theoretical description of the multiple 
features observed. Even the most simple case, that is, the  
dynamics of a hole injected in an antiferromagnetic matrix, remains still largely unsolved. In particular, 
the main disagreement is that in  most of the theoretical calculations --based on the $t-J$ model-- the low energy excitations corresponds to quasiparticle excitations\cite{liu92}, while in the ARPES spectra the width of the low energy peaks is so large that they can not be associated with a physical lifetime of  
quasiparticles\cite{wells95,laughlin97,damascelli03}. In addition to this controversy observed in the cuprates, there has been an increasing interest in the study of the hole motion in different antiferromagnetic backgrounds in order 
to test the validity of the coherent quasiparticle --or spin polaron--  picture. For the unfrustrated $180^{\circ}$ N\'eel order it has been shown, numerically and analytically\cite{kane89,martinez91,liu92,dagotto94,muramatsu00,mishchenko01}, that  the quasiparticle excitations exist for all momenta and for all $J>0$ . On the other hand,  in a highly frustrated case like the kagom\'e lattice, which is believed to be magnetically disordered, completely incoherent spectral functions have been found\cite{poilblanc04}  for all momenta, $J/t=0.4$, and both $t$ signs -- it should be noted that in the square lattice unfrustrated case the particle-hole symmetry leads to the same behavior for both signs of $t$. In addition, we have recently found\cite{trumper04} that in the triangular antiferromagnet, with a $120^{\circ}$ N\'eel order,  an intermediate situation between the square and the kagom\'e geometries arises, since the quasiparticle weight  only vanishes for $t>0$. This result is particularly interesting because it means that the conventional quasiparticle picture can be broken in a semiclassical magnetic background, without invoking spin liquid phases. Besides the mechanism of hole motion assisted by spin fluctuations, already existing in the unfrustrated  case, in the triangular antiferromagnet there appears a free hopping hole
mechanism as a  direct consequence of a ferromagnetic component of the underlying magnetic structure. The latter implies a finite probability of hole motion without emission or absorption of magnons. In our previous work we have shown  that it is the subtle interference between both processes for hole motion that produces the vanishing of the quasiparticle excitations\cite{trumper04}. 

In the present article we will study the hole dynamics in canted antiferromagnetic states where,  by varying the canting angle, it is possible to consider the evolution of the  magnetic background from the antiferromagnetic to the ferromagnetic state. This kind of study, little explored in the literature \cite{vojta98}, allows us to investigate the spectral functions continually from  the AF state --only spin flip assisted processes-- to the ferromagnetic state --only free hopping process--, so as to investigate more carefully how  the non-trivial interference between both hole-motion processes influences the formation of a coherent quasiparticle. To carry out the study we have derived an effective Hamiltonian  from the $t-J$ model using the spinless fermion representation for the kinetic part, and a canted spin wave state for the magnetic part. Within the self consistent Born approximation we have found that, for $J<t$, the hole propagates preferably at two well separated energies: as a coherent spin polaron excitation at low energy, and as a quasi-free hole at higher energy. We were able to associate the former energy scale to the AF component, while the latter to the ferromagnetic one. As the canting angle increases we have observed an important spectral weight transfer from low to higher energy sectors leading to the reduction of the quasiparticle weight. In particular, inside the magnetic Brillouin zone (MBZ), the quasiparticle weight vanishes for $\theta \gtrsim 60^{\circ}$.\\  

This article is organized as follows. In Sec. II we formulate the effective Hamiltonian for a hole moving in a canted AF. In Sec. III we briefly draw the calculation of the hole Green function within the self consistent Born approximation. In Sec. IV we present and discuss the results for the hole spectral functions,  and in Sec. V we state the conclusions of our work.   
\section{Effective Hamiltonian for a hole in a  canted antiferromagnet}

To study the hole motion in a canted antiferromagnet we will study the $t-J$ model in the square lattice\cite{dagotto94}. In order to stabilize a canted phase we add a Zeeman term  that couples only with the spin operators, so that we can vary the canting angle $\theta$ by tuning a {\it fictitious} uniform magnetic field $B$. Actually, such a canted phase might be a consequence of further anisotropic magnetic interactions, but the final effect on the hole dynamics will be the same.
Thus, we use the following $t-J$ Hamiltonian

\begin{eqnarray}
{H}=&H_t+H_J=&-t\sum_{\langle {i},{j} \rangle }(\hat{c}^{\dagger}_{i,\sigma}
                                         \hat{c}_{j,\sigma}+ h.c.)+ \nonumber \\
  & & +J\sum_{\langle {i},{j} \rangle }
{{\bf {S}}}_{i} \cdot {{\bf {S}}}_{j} + B \sum_i S^z_i~,
\label{tj}   
\end{eqnarray}
  
 \noindent where the kinetic part $H_t$ represents the hopping between nearest
neighbors ($<i,j>$) of the square lattice,  with the constraint of no double
occupancy, $\hat{c}_{i,\sigma}=c_{i,\sigma} (1-n_{i,-\sigma}$),
and $H_J$ represents the AF  Heisenberg part along with the Zeeman term. 

\subsection{Magnetic part}

The magnetic part is treated in the spin wave approximation\cite{manousakis91}. It is assumed that the semiclassical  uncanted AF order lies in the $x-y$ plane while the magnetic field $B$ points in the $z$ direction. So, the effect of the  magnetic field is to tilt the spins an angle $\theta$ out of the $x-y$ plane. The generalized Holstein-Primakov transformation for the canted case results 
\begin{eqnarray}
 {\bf S}^x_i&=& \cos \theta \;\; (S-a^{\dagger}_i a_i) -\imath \sqrt{S/2}\;\; 
 \sin \theta \;\;(a_i-a^{\dagger}_i) \nonumber \\
 {\bf S}^y_i&=& \sqrt{S/2} \;\;(a_i+a^{\dagger}_i) \nonumber \\
   {\bf S}^z_i&=& -\sin \theta \;\;(S-a^{\dagger}_i a_i) -\imath \sqrt{S/2}\;\; 
 \cos \theta \;\;(a_i-a^{\dagger}_i), \nonumber 
\end{eqnarray}
\noindent for $i$ $\in$ sublattice $A$, and
\begin{eqnarray}
 {\bf S}^x_j&=& -\cos \theta\;\; (S-b^{\dagger}_j b_j) +\imath \sqrt{S/2} \;\; 
 \sin \theta \;\;(b_j-b^{\dagger}_j) \nonumber \\
 {\bf S}^y_j&=& \sqrt{S/2}\;\; (b_j+b^{\dagger}_j) \nonumber \\
   {\bf S}^z_j&=& \sin \theta\;\; (S-b^{\dagger}_j b_j) + \imath \sqrt{S/2}\;\; 
 \cos \theta \;\;(b_j-b^{\dagger}_j), \nonumber 
\end{eqnarray}
\noindent for $j$ $\in$ sublattice $B$. The Fourier transform of the bosonic operators $a$'s and $b$'s are defined as
$$
a^{\dagger}_i= \sqrt{\frac{2}{N}} \sum_{\bf k} e^{\imath {\bf k} R_i} a^{\dagger}_{\bf k} 
\;\;\;\;\;\\;\; 
b^{\dagger}_j= \sqrt{\frac{2}{N}} \sum_{\bf k} e^{\imath {\bf k} R_j} b^{\dagger}_{\bf k},
$$
\noindent where $N$ is the number of lattice sites and ${\bf k}$ runs along the magnetic Brillouin zone (half of the square lattice Brillouin zone). Up to quadratic order, the magnetic part of the $t-J$ Hamiltonian takes the form  
\begin{eqnarray}
H_J&=& E_C +  \sum_{\bf k}  [  A (a^{\dagger}_{\bf k} a_{\bf k} + b^{\dagger}_{\bf k} b_{\bf k}) + \label{HJ}  \\
&+& D \gamma_{\bf k}   (a^{\dagger}_{\bf k} b_{\bf k} + a_{\bf k} b^{\dagger}_{\bf k})
+  C \gamma_{\bf k} (a_{\bf k} b_{-{\bf k}} + a^{\dagger}_{\bf k} b^{\dagger}_{-{\bf 
k}})] \nonumber 
\end{eqnarray}
\noindent with $A= 4JS \cos2 \theta +B \sin \theta$, $D=JS \sin^2 \theta$, $C=JS \cos^2 \theta$, and 
$$
E_c= -2JS^2N \cos^2\theta-BNS \sin\theta.
$$
\noindent Minimization of $E_c$ with respect to $\theta$ leads to $B=8JS \sin\theta$. This value cancels the linear bosonic terms of the magnetic part and justify the use of the quadratic form Eq. (\ref{HJ}). Furthermore, if this value of $B$ is replaced in $E_c$ and $A$, $H_J$ can be written  as function of the canting angle $\theta$ only. If $H_J$ is expressed in a matrix form the dynamical matrix has a  dimension $4 \times 4$. The diagonalization of $H_J$ is performed in two steps: first, a transformation $\alpha^{\dagger}_{\bf k} = a^{\dagger}_{\bf k}+b^{\dagger}_{\bf k}  $ and   
$\beta^{\dagger}_{\bf k} = a^{\dagger}_{\bf k}-b^{\dagger}_{\bf k} $ that renders the dynamical matrix diagonal by  blocks of $2\times2$ and, then, a Bogoliubov transformation from $\alpha, \beta$ to the new bosonic operators $\eta, \nu$
\begin{eqnarray}
\alpha_{\bf k}&=&u^{+}_{\bf k} \eta_{\bf k} + v^{+}_{\bf k} \eta^{\dagger}_{-{\bf k}} \nonumber \\
  \beta_{\bf k}&=&u^{-}_{\bf k} \nu_{\bf k} + v^{-}_{\bf k} \nu^{\dagger}_{-{\bf k}} \nonumber \end{eqnarray}
 with  Bogoliubov coefficients,
 $$u^{\pm}_{\bf k}=  \left(\frac{A \pm D\gamma_{\bf k} } {2\omega^{\pm}_{\bf k}} +\frac{1}{2} \right)^{1/2}$$
and 
$$
 v^{\pm}_{\bf k}= \mp \frac{\gamma_{\bf k}}{|\gamma_{\bf k}|} \left(\frac{A \pm D\gamma_{\bf k} }{ 2\omega^{\pm}_{\bf k} } -\frac{1}{2}\right)^{1/2}.$$  
  Once these transformations are taken into account the harmonic magnetic part results
 \begin{eqnarray}
 H_J= E_c-2JSN+ \sum_{\bf k}& \left[ \right.(\omega^+_{\bf k}& +\omega^{-}_{\bf k})+  \nonumber \\
 &  + \omega^+_{\bf k}& \eta^{\dagger}_{\bf k} \eta_{\bf k} + \omega^{-} \nu^{\dagger}_{\bf k}\nu_{\bf k} \left. \right] \nonumber
 \end{eqnarray}
 \noindent with  the two magnon dispersion branches  
 
 \begin{equation}
 \omega^{\pm}_{\bf k}= 2JS \sqrt{(1 \pm \gamma_{\bf k}) (1 \mp \cos \theta \gamma_{\bf k} )}, \nonumber
 \end{equation}
  
\noindent defined in the magnetic Brillouin zone and $\gamma_{\bf k}= (\cos k_x+\cos k_y)/2$. As ${\bf k}\to(0,0)$, the dispersion $\omega^{-}_{\bf k}\rightarrow 0$  linearly, while 
 $\omega^{+}_{\bf k}\rightarrow4JS \sin\theta$ quadratically. For this reason, hereafter,
  $\omega^-_{\bf k}$ and $\omega^{+}_{\bf k}$ will be called  
   the AF and ferromagnetic bands, respectively. Furthermore, since $\omega^{+}_{{\bf
k}+ (\pi,\pi)}= \omega^{-}_{\bf k}$, it can be seen that in the one magnon operator description, where the translational symmetry of the square lattice is not broken, it is recovered the same dispersion of ref. \cite{zhitomirsky98} if $\omega^{+}_{\bf k}$ is unfolded to the square Brillouin zone.       

\subsection{Kinetic part}
 In order to fulfill the basic requirement of the $t-J$ model, the no double occupancy constraint, we use  the following spinless fermion transformation\cite{liu92} for the canted case
\begin{eqnarray}
\hat{c}_{i\uparrow}&=& \cos \frac{\theta}{2} g_{i} +\sin\frac{\theta}{2}  g^{\dagger}_{i} a_i \nonumber \\
\hat{c}_{i\downarrow}&=& \cos \frac{\theta}{2} g^{\dagger}_{i} a_i - \sin \frac{\theta}{2} g_i, 
\label{SA}
\end{eqnarray}
for $i$ $\in$ sublattice A
\begin{eqnarray}
\hat{c}_{j\uparrow}&=& \sin \frac{\theta}{2} f_{j} +\cos\frac{\theta}{2}  f^{\dagger}_{j} b_j \nonumber \\
\hat{c}_{j\downarrow}&=& \sin \frac{\theta}{2} f^{\dagger}_{j} b_j - \cos \frac{\theta}{2} f_j, 
\label{SB}
\end{eqnarray}
for $j$ $\in$ sublattice B,
where $g_{i}, f_j$ are the fermionic hole operators and $a_i, b_j$ are the Holstein Primakov bosons. Replacing (\ref{SA}) and (\ref{SB}) in the kinetic part of eq.(\ref{tj}), and retaining terms up to third order it results
\begin{eqnarray}
H_t=-t \;\;\sin \theta \sum \;\; \;\; \gamma_{\bf k} (g_{\bf k} f^{\dagger}_{\bf k} +\;\;f_{\bf k} g^{\dagger}_{\bf k} )-  \nonumber \\
- t \sqrt{\frac{2}{N}} \cos \theta \sum_{{\bf k}{\bf k}^{'}} [ \gamma_{\bf k}
\;\;g_{\bf k} f^{\dagger}_{\bf k'} b_{{\bf k'}-{\bf k}} - \nonumber \\
 -\gamma_{\bf k'} \;\; g_{\bf k} f^{\dagger}_{\bf k'} a^{\dagger}_{{\bf k}-{\bf k'}} + h.c.]  .
\label{qubic}
\end{eqnarray}

\noindent    Now, it is convenient to define the new fermionic bonding and antibonding  operators
 
 \begin{equation}
 g_{\bf k}= \frac{l_{\bf k}+m_{\bf k}}{\sqrt{2}} \;\;\;\;\;f_{\bf k}= \frac{l_{\bf k}-m_{\bf k}}{\sqrt{2}},
 \label{newfermion}
 \end{equation}

\noindent respectively. When fermions $g$ and $f$ are expressed as a function of the bonding and antibonding fermions (\ref{newfermion}), and the bosons $a$ and $b$ in terms of the Bogoliubov operators $\eta,\; \nu$, the kinetic part (\ref{qubic})
can be rewritten as 

\begin{eqnarray}
H_t=\sum_{\bf k} \epsilon_{\bf k} (m^{\dagger}_{\bf k}m_{\bf k}- l^{\dagger}_{\bf k}l_{\bf k})+ \nonumber \\
+ \sqrt{\frac{2}{N}} \sum_{{\bf k}{\bf q}} \left\{ [M^{+}_{{\bf q}{\bf k}}\; \eta^{\dagger}_{{\bf k}-{\bf q}}
\;(m_{\bf k} l^{\dagger}_{\bf q}- l_{\bf k} m^{\dagger}_{\bf q}) +\;\;h.c.\;] + \right. \nonumber \\
\left.+ [M^{-}_{{\bf q}{\bf k}} \; \nu^{\dagger}_{{\bf k}-{\bf q}}
\;(l_{\bf k} l^{\dagger}_{\bf q}- m_{\bf k} m^{\dagger}_{\bf q}) +\;\;h.c.\;] \right\},
\label{Htfinal}
\end{eqnarray}

\noindent with the free hopping band $\epsilon_{\bf k}=4t\sin\theta\; \gamma_{\bf k}$
and the vertex interactions
\begin{equation}
M^{\pm}_{{\bf q}{\bf k}}= -\frac{t}{2} \cos \theta \;\; (u^{\pm}_{{\bf k}-{\bf q}} \gamma_{\bf q} 
-v^{\pm}_{{\bf q}-{\bf k}} \gamma_{\bf k}).
\end{equation}
 
\noindent There are two mechanisms for charge motion. The first one is a free hopping process, first line of eq. (\ref{Htfinal}), that naturally appears with the canting of the AF order. The second one represents a hopping processes magnon assisted by the ferromagnetic and the antiferromagnetic bands, respectively.
When the canting angle $\theta=0^{\circ}$, the free hopping term vanishes and two degenerate antiferromagnetic bands are recovered. So, as it is expected for the case of hole motion in a pure AF matrix,  the kinetic part is only described by a hole coupled to AF magnons\cite{martinez91,liu92}. On the other hand, when the canting angle is such that the underlying magnetic order is ferromagnetic, $\theta=90^{\circ},$ the magnon assisted hopping  disappears and the only  mechanism available for hole motion is the free hopping term.      
Therefore, it is possible to interpolate the hole motion continually between the pure AF state and the ferromagnetic state, so as to investigate carefully how  the non-trivial interference between both hole-motion processes influences the formation of a coherent quasiparticle.

\section{The self-consistent Born approximation}
   
\begin{figure}[h]
\vspace*{0.cm}
\includegraphics*[width=0.35\textwidth]{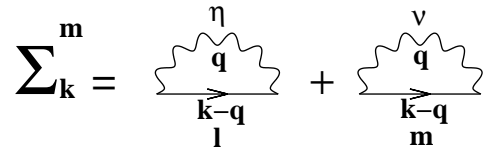}
\caption{Relevant contributions in the SCBA to the self energy of the Green function corresponding  to the antibonding fermion $m$. The wiggly lines represents free magnon Green functions for $\nu$ and $\eta$, while the straight lines represents the dressed fermionic Green functions $G^{m}$ and $G^{l}$. For the self energy of the bonding fermion $l$, $m$ and $l$ must be interchanged.}
\label{fig1}
\end{figure}

The use of the two magnetic sublattices requires the definition of the  two Green functions (see eq.(\ref{Htfinal})) 
 
 $$
 G^{m}_{\bf k}(\omega)= \frac{1}{\omega - \epsilon_{\bf k} - \Sigma^{m}_{\bf 
 k}(\omega)}\;\; ; \;\;\; G^{l}_{\bf k}(\omega)= \frac{1}{\omega + \epsilon_{\bf k} - \Sigma^{l}_{\bf 
 k}(\omega)} 
$$

 \noindent along the magnetic Brillouin zone.  
Taking into account the interacting terms of the Hamiltonian (\ref{Htfinal}) within the self consistent Born approximation, it is straightforward to see that 
 there are two contributions to each self-energy (see for instance  Fig. \ref{fig1} for the self  energy of the antibonding fermion  $m_{\bf k}$). 
 
 \noindent A standard procedure leads to two coupled  self-consistent equations for the self energies, 

 \begin{eqnarray}
 \Sigma^{m (l)}_{\bf k}(\omega)=  \sum_{\bf q} \left\{|M^{+}_{{\bf k}+{\bf q}\;{\bf k}}|^2
 G^{l(m)}_{{\bf k}+{\bf q}}(\omega-\omega^{+}_{\bf q}) + \right. \nonumber \\
\left. +|M^{-}_{{\bf k}+{\bf q}\;{\bf k}}|^2
 G^{m(l)}_{{\bf k}+{\bf q}}(\omega-\omega^{-}_{\bf q}) \right\}, 
 \label{self}
 \end{eqnarray}
 
  \noindent that will be solved numerically.
 At this point, it is enlightening to relate the bonding $G^{m}_{\bf k}(\omega)$ and 
 antibonding $G^{l}_{\bf k}(\omega)$, defined in the magnetic Brillouin zone, with the more physical hole Green function $G^{h}_{\bf k}(\omega)$ defined in the whole Brillouin zone. The hole operator $h$ is defined as $h_i= f_i$ for $i$ $\in$ sublattice A and $h_i= g_i$ for $i$ $\in$ sublattice B. Then, if $h_{\bf k}$ is splitted as
 $$
 h_{\bf k}=\frac{1}{N} \sum_{i\epsilon A} f_i e^{\imath {\bf k} {\bf R_i}}
 +\frac{1}{N} \sum_{i\epsilon B} g_i e^{\imath {\bf k} {\bf R_i}},
 $$
\noindent it is straightforward to find the operatorial relations
  $ h_{\bf k}=\frac{1}{\sqrt{2}} (f_{\bf k} + g_{\bf k})= l_{\bf k} $
for ${\bf k}$ inside the magnetic Brillouin zone and $
 h_{{\bf k}}=\frac{1}{\sqrt{2}} (f_{{\bf k}+{\bf Q}} - g_{{\bf k}+{\bf Q}})= m_{{\bf k}+{\bf Q}}
 $
for ${\bf k}$ outside the magnetic Brillouin zone. These relations between the $h_{\bf k}, l_{\bf k}$ and $m_{\bf k}$, imply the following relation between the Green functions

\begin{eqnarray}
G^h_{\bf k} &=& G^l_{\bf k}  \;\;\; \textrm{inside} \;\;\;\textrm{MBZ} \nonumber \\
G^h_{\bf k} &=& G^m_{{\bf k}+{\bf Q}} \;\;\; \textrm{outside}\;\;\; \textrm{MBZ}.
\label{greenrel}
\end{eqnarray}
   
\section{Results}
\subsection{Spectral function}
We have solved numerically the self consistent equations (\ref{self}) for $\Sigma^{m(l)}_{\bf k}(\omega)$, using cluster sizes up to  $N=40 \times 40$ and a frequency grid of $20000$ points. Then, we have calculated the corresponding spectral function $A^{h}_{\bf k}(\omega ) = - \frac{1}{\pi}ImG_{\bf k}^{h}(\omega)$ for several canting angles. 

\begin{figure}[h]
\vspace*{0.cm}
\includegraphics*[width=0.4\textwidth]{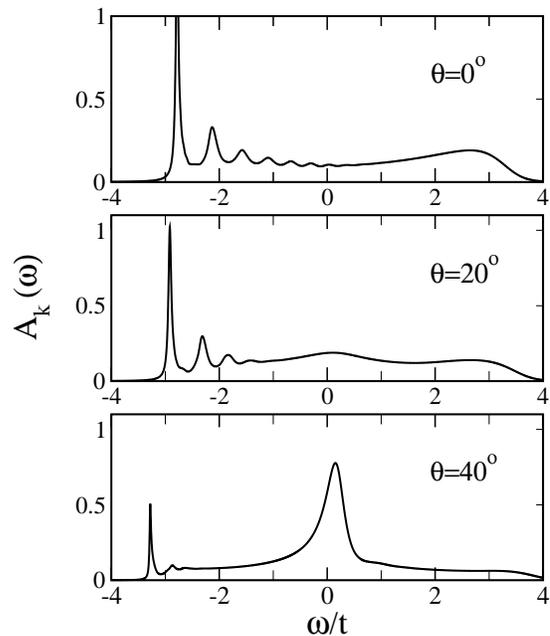}
\caption{ Spectral function for ${\bf k}=(\pi/2, \pi/2)$ and $J/t=0.1$.}
\label{fig2}
\end{figure}

In Fig. \ref{fig2} it is shown the spectra for ${\bf k}=(\pi/2,\pi/2)$ and $J/t=0.1$. We have chosen these particular  momentum and coupling regime since for them, the main features of the spectra are clearly differentiated, and therefore it is easier to identify  the underlying mechanisms for hole motion. The case $\theta=0^{\circ}$ (upper panel) corresponds to the spectral function of a hole in a pure AF matrix. This result has been already obtained by one of us \cite{martinez91} and others authors \cite{liu92} using the SCBA. The spectra extend over a frequency range  $\sim 8t$ with a low energy sector composed by a delta peak at the bottom  of the spectra along with  several resonances of finite lifetime above it. The former is associated with a quasiparticle excitation (QP), whose bandwith is of order $J$; whereas the latter resonances can be identified with
string excitations since its energies scale as $E_{string}\sim (J/t)^{2/3}$. On the other hand, there is  an incoherent part corresponding to the shoulder located at an energy $\sim 3 t$. \\

The low and high energy structure of the spectra can be traced back  to the coupling between the hole and  
 the underlying AF order.  
 {\it Low energy sector:} as the hole moves, with a characteristic time of order $1/t$, the AF order is locally disturbed   
 leaving a string of overturned spins. Meanwhile, the zero 
 point spin fluctuations above the classical N\'eel state $|N>$,  contained in the quantum AF ground state $|AF>= exp(-\sum_{ij} u_{ij} a^{\dagger}_i b^{\dagger}_j)|N>,$ \cite{barentzen98} repair pairs of frustrated spins at a characteristic time of order $1/J$. 
It is clear that, in the weak coupling regime ($J>t$),  the magnetic string of overturned spins can be completely erased  by  the zero point spin fluctuations, and  the hole surrounded by an AF cloud emerges as a coherent quasiparticle excitation. However, in the strong coupling regime ($J<t$), the magnetic string is only partially erased and,  besides the low energy QP excitation, there are higher energy processes --strings excitations-- corresponding to the hole inside a linear potential generated by the overturned spins. This picture has been widely confirmed by several numerical and analytical techniques\cite{martinez91,liu92,dagotto94,muramatsu00,mishchenko01}. {\it High energy sector:} to describe the incoherent part of the high energy sector, it is convenient to take a closer look at the zero point spin fluctuations. In particular, if the exponential function in $|AF>$ is developed in a Taylor series, the quantum AF ground state can be written as 
$|AF> = |N>+|fluct>$, where the fluctuations  can be seen as a sum of $S^z$ conserving  terms like $ a^{\dagger}_{i_{1}} a^{\dagger}_{i_{2}}...a^{\dagger}_{i_{n}}  b^{\dagger}_{j_{1}} b^{\dagger}_{j_{2}}....b^{\dagger}_{j_{n}}|N>$. As we have stated before, these fluctuations erase part of the strings, but they also generate small ferromagnetic clusters wherein the hole can propagate freely. This explains the broad shoulder centered at $\omega \sim 3 t,$ that is, the finite probability to find the hole propagating at considerable high energies above the QP excitation. In fact, we have extended our calculation to the anisotropic Heisenberg model, and we have 
effectively  found a suppression of the shoulder  as well as an enhancement of the string resonances as the Ising limit is approached.\\
 
Now we discuss the evolution of the spectra with the canting angle. At $\theta=20^{\circ}$ (middle panel of Fig. \ref{fig2}), there appears a  classical ferromagnetic component in the underlying magnetic order, while the zero point spin fluctuations  get reduced. This adds a free hopping mechanism for the hole motion, represented by the tight binding term of eq. (\ref{Htfinal}). Such an additional mechanism competes with the magnon-assisted, as well as with the incoherent hopping processes driven by the zero point spin fluctuations, resulting in a spectral weight transfer from the low energy sector and the incoherent shoulder  to an energy located at $\omega \sim 0.$ For $\theta = 40^{\circ},$ (lower panel of Fig. \ref{fig2}) most of the spectral weight is dominated by this new mechanism, signaled by a finite lifetime resonance, $t$-resonance, located at an energy close to  $\epsilon_{\bf k}=\;-t \sin\;\theta \;\gamma_{\bf k}$. So that, as the canting angle increases, the probability of finding the hole moving freely along the classical ferromagnetic channel becomes more important than both, the magnon-assisted  and the incoherent hopping processes driven by zero point spin fluctuations. A similar scenario of a low energy excitation  coexisting with a higher energy long-lived  resonance, dispersing as a free band, has been found  in recent high-resolution photoemission spectra from the insulating cuprates $Ca_2CuO_2Cl_2$\cite{ronning05}. While in the cuprate the appearance of the high energy part of the spectra has been ascribed to hoppings to first and further neighbors, in our present calculation the high energy part, dispersing like a free band, is obtained with the canting angle.\\      

\begin{figure}[h]
\vspace*{0.cm}
\includegraphics*[width=0.4\textwidth]{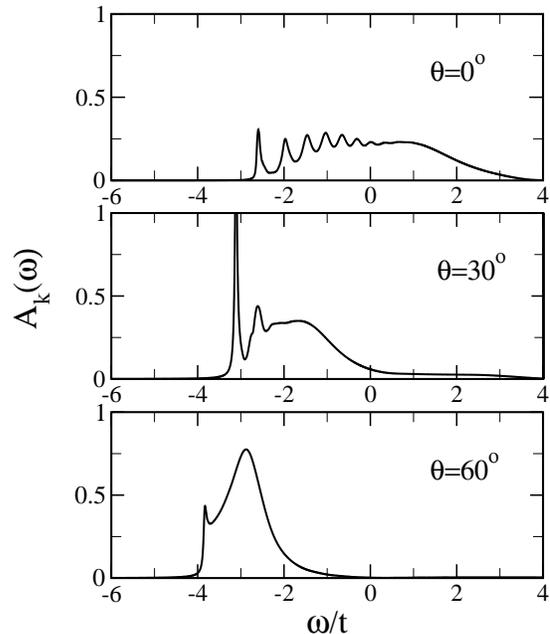}
\caption{ Spectral function for $\bf k=(0.8\pi,0.8\pi)$ and $J/t=0.1$.}
\label{fig3}
\end{figure}

The evolution of the spectral function with $\theta$ becomes more complex when the ${\bf k}$ dependence is taken into account. In particular, for ${\bf k}$ outside the magnetic Brillouin zone  the QP and string excitations start to overlap with the $t$-resonance at the low energy sector (see Fig \ref{fig3}), while for ${\bf k}$ inside the magnetic Brillouin zone something similar to ${\bf k}=(\pi/2,\pi/2)$ in Fig. \ref{fig2} occurs, namely, the peaks are quite separated. As the canting angle is increased, there is a transfer of spectral weight from the magnon assisted to the free hopping process as a consequence of the reduction of the vertex interaction with $\theta$, $M^{\pm}_{{\bf q}{\bf k}}\sim \cos\theta$. 
It is worth to stress that the low energy QP excitation has its origin in the coherent 
scattering between the hole and the magnons, so the reduction of the scattering rate 
renders the magnon-assisted hopping process less effective than the free hopping one, 
that is originated by the ferromagnetic component of the underlying magnetic structure. 
Finally, for $\theta=90^{\circ}$, the only allowed hole motion process is the free one along the completely ferromagnetic order, thus, the whole spectral function becomes a single delta peak. 

\subsection{Quasiparticle excitations}
\begin{figure}[h]
\vspace*{0.cm}
\includegraphics*[width=0.4\textwidth]{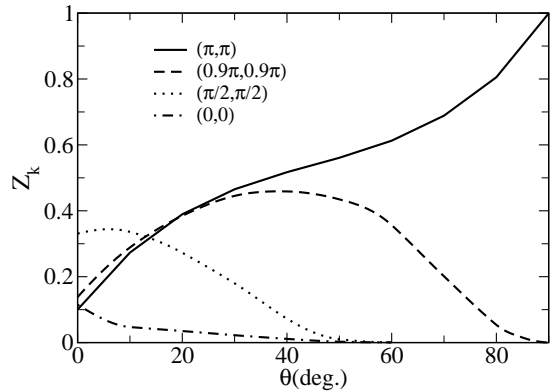}
\caption{QP weight as a function of the canting angle $\theta$ for several momenta and $J/t=0.4$.}
\label{fig4}
\end{figure}

In this section we analyze the quasiparticle excitations in the low energy sector of the spectra. This can be quantified by the QP weight  $z_{\bf k}=|<\Phi_{\bf k}|h^{\dagger}_{\bf k}|AF>|^2$ that gives a measure of the overlap between the state of a bare hole created on the AF background and the quasiparticle state $|\Phi_{\bf k}>$. In our case we computed the QP weight using the well known relation $z_{\bf k}= (1- \partial \Sigma_{\bf k}(\omega)/\partial \omega|_{E_{\bf k}})^{-1}$.  
It is worth to note that  the complex ${\bf k}$ dependence of the spectral weight transfer with $\theta$, mentioned in the previous section, is clearly manifested in the QP excitations. In Fig. \ref{fig4} it is shown the QP weight versus $\theta$ for $J/t=0.4$ and several momenta. 
On one hand,  for ${\bf k}$ inside the magnetic Brillouin zone  the effect of the canting is to monotonically decrease the QP weight until it vanishes at around $\theta \sim 60^{\circ}$ (see for instance, ${\bf k=(\pi/2,\pi/2)}$ and ${\bf k}=(0,0)$ of Fig. \ref{fig4}). In this region the spectra is characterized by the two well separated peaks (see $\theta=40^{\circ}$ in Fig. \ref{fig2}) where the QP excitations have a magnon assisted origin. So, the vanishing of the QP weight with $\theta$ is due to the reduction of the vertex interaction with $\theta.$
On the other hand, for ${\bf k}$ outside the magnetic Brillouin zone --$k=(\pi,\pi)$ and $k=(0.8\pi,0.8\pi)$-- initially the QP weight increases with $\theta.$ This is an unexpected behavior if the $\theta$ dependence of the vertex interaction is taken into account again.  However, in this sector of the Brillouin zone, the magnon assisted and the free hopping processes merge at the low energy sector increasing the QP weight (see $\theta=30^{\circ}$ in Fig. \ref{fig3} ). This can be seen as a constructive interference between the hole motion processes in the formation of the QP. For greater angles the QP weight is dominated by the vertex interaction and goes to zero as $\theta\rightarrow90^{\circ}$. At $\theta=90^{\circ}$ the  magnon assisted process vanishes for all ${\bf k}$ and the only allowed process is the free hole motion along the ferromagnetic channel, jumping  $z_{\bf k}$ from zero to unity. ${\bf k}=(\pi,\pi)$ is a unique case where the QP weight increases monotonically to unity due to the strict energy coincidence of both hole motion processes. For this case we can say that there  is always a constructive interference of the processes. It should be noted that the QP ground state momentum evolves with the canting angle along the diagonal $(\pi/2,\pi/2)\to(\pi,\pi)$, being $(\pi,\pi)$ the ground state momentum for $\theta$ greater than $40^{\circ}$ when $J/t=0.4$. 
 
The advantage of using explicitly two sublattices in our calculation is the possibility to analyze separately the coupling of the hole with the ferro and AF magnons. For instance, if we cancel 
$M^+$ ($M^-$) in the kinetic part (\ref{Htfinal}), the coupling of the hole with the ferro (AF) band is omitted. Under this condition the self-consistent equations  (\ref{self}) for $\Sigma^{m(l)}_{\bf k}(\omega)$ can be solved and, via the relations (\ref{greenrel}), it is obtained $G^h_{\bf k}(\omega)$ without the effect of the ferro or the AF excitations on the hole motion.

\begin{figure}[t]
\vspace*{0.cm}
\includegraphics*[width=0.4\textwidth]{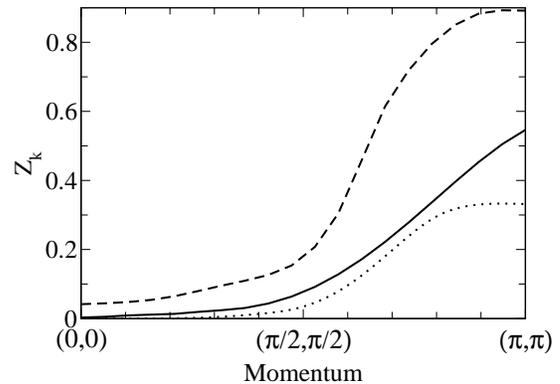}
\caption{ QP weight  along $(0,0) \to (\pi,\pi)$ for $J/t=0.4$ and $\theta=40^{\circ}$. Contribution from  the ferromagnetic band (dashed line), AF band (dotted), and both (solid line) in the SCBA.   }
\label{fig5}
\end{figure}

For instance, in Fig. \ref{fig5} it is shown separately the ferro (dashed line) and the AF (dotted line) contributions to the QP weight, along with the complete prediction (solid line), for $J/t=0.4$ and $\theta=40^{\circ}$. It is observed that the QP weight is greater for the ferromagnetic than for the AF contribution. Furthermore, when both magnetic bands are considered, the QP weight resembles that of the AF character. 
This can be seen as a consequence of the different momentum dependence of the interaction vertices $M^+, M^-.$  For all ${\bf k}$ and ${\bf q}\sim 0$, the vertex $M^{-}_{{\bf q}{\bf k}}\sim \sqrt{q},$ while $M^{+}_{{\bf q}{\bf k}} \sim const+q$. 
As stated before, since the QP excitation has a magnon assisted character driven 
by the vertex interaction, a stronger coupling produces an enhancement of the 
QP weight. So that, the coupling of the hole with the ferromagnetic band is more 
coherent than with the AF one, and when both couplings are considered together 
the QP weight follows the less coherent AF contribution.    

Now we discuss the $J/t$ dependence of the QP excitations. 
As $J/t$ increases, we have observed that the 
character of the QP excitations changes from a many body state resulting from the dynamical coherent coupling of the hole with the magnons, to a free hole state 
weakly perturbed by the magnons. This crossover can be seen in more detail analyzing
the QP wavefunction $|\Psi_{\bf k}>$\cite{reiter94,ramsak98,trumper04}. In general,  $|\Psi_{\bf k}>$ can be expressed 
as a sum of terms with one hole and different number of magnons, that in our case can be written as
$$
|\Psi_{\bf k}>=a^{(0)}_{\bf k}h^{\dagger}_{\bf k}|AF>+
\sum_{{\bf q_1},\sigma}a^{(1\sigma)}_{\bf k,\bf q_1}h^{\dagger}_{\bf k-q_1}{\xi^{(\sigma)}}^{\dagger}_{\bf q_1}|AF>+\cdots,
$$  
where $\sigma=\pm$, and $\xi^{(\sigma)}$ represents the Bogoliubov operators $\xi^{(+)}=\eta$ and $\xi^{(-)}=\nu.$  As $J/t$ increases, the multimagnon processes are energetically
more expensive and their contributions to the QP wavefunction are notably reduced, whereas the zero and the one magnon terms become the relevant ones\cite{trumper04}. 
Within the SCBA\cite{ramsak98} the one magnon coefficient is $a^{(1\sigma)}_{\bf k,\bf q_1}=z_{\bf k}M^{\sigma}_{{\bf k}{\bf q_1}}G^h_{{\bf k}-{\bf q}_1}(E_{\bf k}-\omega^{\sigma}_{{\bf q}_1}),$ while the zero magnon coefficient is $a^{(0)}_{\bf k}=
z_{\bf k}.$ The many-body state character of the QP excitation is signaled by the dependence of $a^{(1\sigma)}$ with the hole Green function, which carries the information of the dynamical coupling of the hole with the magnons. In the weak coupling regime, large values of $J/t,$ as $z_{\bf k} \to 1$ and $E_{\bf k} \to
\epsilon_{\bf k}$ (see below), $a^{(1\sigma)}$ becomes the first-order coefficient of a conventional Rayleigh-Schr\"odinger perturbation theory,
$a^{(1\sigma)}_{\bf k q}=M^{\sigma}_{{\bf k},{\bf q}}/(\epsilon_{\bf k}-\epsilon_{{\bf k}-{\bf q}}-\omega^{\sigma}_{\bf q}).$ In this coupling regime, the character of the QP wavefunction is that of a free hole state weakly renormalized by the one magnon excitations. This state correspond to the above mentioned $t$-resonance.   

\begin{figure}[h]
\vspace*{0.cm}
\includegraphics*[width=0.4\textwidth]{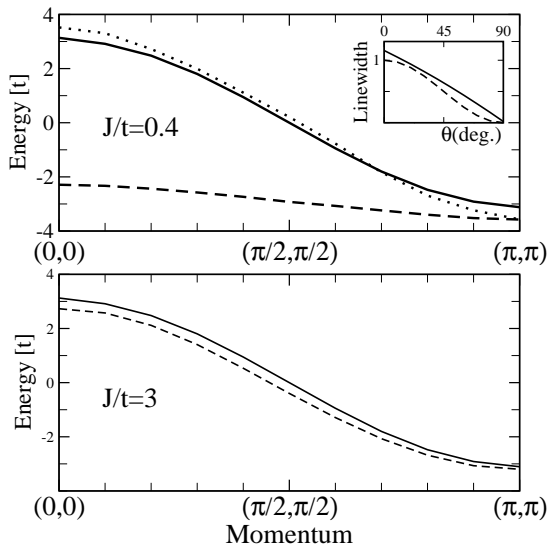}
\caption{Energy dispersion of the QP excitation (dashed line), bare hole excitation (solid line), and {\it t} resonance excitation (dotted line) for an angle $\theta=50^{\circ}$. Top panel is for    $J/t=0.4$. Inset: linewidth of the {\it t} resonance (solid line) and $\cos^2 \theta$ (dashed line) as a function of $\theta$. Bottom panel is for $J/t=3$. For this regime the QP and the {\it t}-resonance excitations are the same.}
\label{fig6}
\end{figure}

In Fig. \ref{fig6} we show the QP, the $t$-resonance, and the bare hole energy dispersions for a canting angle $\theta=50^{\circ}$, in the strong, $J/t=0.4,$ and the weak, $J/t=3$ coupling regimes. 
On one hand, for strong coupling (top panel), the QP energies are well separated from the bare hole and the $t$-resonance energies, indicating the highly non-perturbative character of the hole motion assisted by the magnons. On the other hand, once the crossover took place, for weak coupling (bottom panel), the QP and the $t$-resonance excitations have merged into a weakly perturbed state, whose dispersion closely follows the bare hole energy one. We have verified that the $t$-resonance dispersion is very well approximated by the weak coupling expression 
$$E^{t}_{\bf k}=\epsilon_{\bf k}+\sum_{\sigma \bf q}\frac{|M^{\sigma}_{\bf k,q}|^2}{\epsilon_{\bf k}-\epsilon_{\bf k-q}-\omega^{\sigma}_{\bf q}},$$
in both, the strong and the weak coupling regimes. So, we can assure that the 
$t$-resonance can always be identified with the bare hole propagating along the ferromagnetic component, weakly perturbed by magnons. 
Another indication of the perturbative character of the $t$-resonance 
is the scaling of its linewidth with $M^2 \sim \cos^2\theta$ (see inset of Fig. \ref{fig6}). It is worth to stress that, while the decrease of the vertex interaction $M$ renders the $t$-resonance more coherent due to its perturbative
character, it suppresses the coherence of the non-perturbative magnon assisted process.  
 
Finally, we have found that, in the weak coupling, the QP weight is
well approximated by the expression\cite{kane89}
$$z_{\bf k}=\frac{1}{1+\sum_{{\bf q}\sigma} \left(M^{\sigma}_{{\bf k}{\bf q}}/\epsilon_{\bf k}-\epsilon_{{\bf k}-{\bf q}}-\omega_{\bf q}\right)^2}\to 1.$$

\subsection{Strings excitations}
In this section we analyze the dependence of the strings with the canting angle. Our general picture is based on Fig. \ref{fig2}. For $\theta=0^{\circ}$ there are several resonances above the QP peak which are interpreted as string excitations that results from the linear potential generated by paths of overturned spins in the AF background. This was confirmed in the SCBA\cite{liu92} by noting that their energies
scale as $E_{string}\sim (J/t)^{2/3}$. Notice that in the string picture the same exponent $2/3$ is obtained for the QP energy since the precursors of the quasiparticle excitation are just the strings\cite{dagotto94}. As the ferromagnetic channel of the underlying magnetic background is enhanced, as a consequence of greater canting angles, the AF channel is weakened and the string excitations start to  smear gradually until at angles around $\theta=40^{\circ}$ they disappear. This can be related to the fact that, at higher angles $\theta,$ the AF channel has been reduced, so that the hole feels a less (sublinear) confining potential. To quantify this behavior we have computed the energy exponents of the QP  and first string energy versus $(J/t)^{\alpha}$ 
by varying the canting angle, for a range of $0.01<J/t<0.3$. In general, for all the momenta investigated, we have found values of $\alpha$ quite close to $2/3$ for the QP and the first string excitation energies. As $\theta$ is increased the value of $\alpha$ remains approximately constant until for for angles greater than $40^{\circ}$  there is a depart from $2/3$ to larger values. In Table I we display the dependence of the $\alpha$ exponents for several angles for the QP and the first string energies, at 
momentum ${\bf k}=(\pi/2,\pi/2)$. 
\begin{table}
\caption{Exponents $\alpha$ for the energy scaling  $E\sim (J/t)^{\alpha}$ of the QP and the first string calculated for a lattice of $N=40\;\times\;40$ and several angles for ${\bf k}=(\pi/2,\pi/2)$. In the case of QP energies the parameter region is $0.01<J/t<0.3$ while for the strings $0.01<J/t<0.1$. To obtain better fits it has been considered the energy contribution of the magnetic background as in ref. \cite{martinez91}.}
\begin{tabular}{cc|cccc|cccc|cccc|ccc|ccc|cc}
\hline
\hline
 
         $\theta$  &   & &      $0^{\circ}$   & & && $10^{\circ}$ &&  &  &  $20^{\circ}$    &&&& $30^{\circ}$ &&&$40^{\circ}$ &&& $50^{\circ}$ \\
\hline
$QP$       && &0.66   &&&&   0.66 &&&& 0.67  &&&&0.69&&&0.72&&&0.74 \\
$String$   && &0.68   &&&&  0.68 &&&&  0.65 &&&& 0.65&&& 0.67&&&0.74 \\
\hline
\end{tabular}
\label{table1}
\end{table}   
To provide a deeper insight of the type of potential implied by these energy exponents $\alpha,$ it is useful to make a simple variational calculation of one particle within a one-dimensional potential. Let us assume a Hamiltonian $H=T+V$ where $T$ is the kinetic energy and the potential energy is of the type $V(x)= (J/t) |x|^{\beta}$. Then, if we propose a variational wave function for the particle of Gaussian form $\phi(x)= A \exp{-\frac{K^2x^2}{2}}$, with $K$ the variational parameter, it is straightforward to  calculate the energy $E(K)=\frac{<\phi|H|\phi>}{<\phi|\phi>}$. Subsequent minimization with respect to $K$ leads to an energy dependence $E\sim (J/t)^{2/(\beta+2)}.$ For a linear potential it is recovered the expected value $\alpha=2/3$, whereas   a  sublineal potential, $\beta<1$, implies $\alpha > 2/3$, in agreement with table I.

\section{Conclusions}           
We have made a detailed analysis of the competing mechanisms for  hole motion  in a canted antiferromagnetic background. To study the hole dynamics we have introduced the spinless fermion representation for the constrained fermions of the $t-J$ model on a square lattice. We have  modeled the canted antiferromagnetic background adding  a Zeeman term to the $t-J$ model, whose effect is to tilt the N\'eel order giving rise to a ferromagnetic component. The problem thus formulated allows to study the hole dynamics 
continually from the pure antiferromagnetic case to the pure ferromagnetic one. As it is well known, in an unfrustrated N\'eel order a hole can only propagate by emitting and absorbing magnons, while for the pure ferromagnetic case the hole propagates freely. Here we have analyzed the evolution of the hole dynamics as a function of the  canting angles by computing the hole spectral function. For this purpose we have use a reliable analytical method for the single hole case like the self consistent Born approximation.

We have found a complex momentum and canting angle dependence of the spectra. 
For $t>J$, the hole propagates preferably at two well separated energies: as a coherent spin polaron excitation at low energy, and as a quasi-free hole at higher energy. In particular, from moderate to large canting angles, the quasiparticle spectral weight is considerably reduced, vanishing for momenta inside the magnetic Brillouin zone. This unexpected result is a consequence of the interference of the two mechanism for hole motion, originated from the antiferromagnetic and the ferromagnetic components of the underlying magnetic order. In the strong coupling regime ($t>J$) the quasiparticle excitation has its origin in the AF component, namely, it is a many body state composed by a hole coherently coupled with the magnons. For this reason, when the canting angle increases, so AF component of the magnetic order  is reduced, the QP weight goes to zero. Closely related to the QP excitations, we have found string excitations that rapidly smear out as the canting angle is increased. On the other hand, at higher energies we have found rather long-lived resonances related to the motion of the hole along the ferromagnetic component. As the ferromagnetic component increases with the canting angle, these resonances become more pronounced. Even in the strong coupling regime, we were able to fit the position and the linewidth of these resonances by a conventional perturbative calculation. As the system moves to the weak coupling regime we found that the character of QP excitations undergoes a crossover from the  many-body spin polaron to  a free hole  state weakly perturbed by magnons.      

We would like to emphasize that these features, obtained for the hole motion in a canted antiferromagnet, are generic and they could also be observed in other magnetic systems, where the anisotropies of the magnetic interactions lead to a canted magnetic state like Dzyaloshinsky-Moriya  or Ising anisotropies. \\

\acknowledgements The authors acknowledge the Brasil-Argentina Scientific Agreement
through project CAPES/SECYT 088/05. This work was also supported by the ANPCYT under grant PICT2004 N$\circ$ 25724. I. J. H. thanks fundaci\'on J. Prats for partial support.

\end{document}